# Nonreciprocal transport in a bilayer of MnBi$_2$Te$_4$ and Pt

Chen Ye[1,†], Xiangnan Xie[2,†], Wenxing Lv[3,†], Ke Huang[1], Allen Jian Yang[1], Sicong Jiang[4,5], Xue Liu[1,6], Dapeng Zhu[7,8], Xuepeng Qiu[9], Mingyu Tong[10], Tong Zhou[2], Chuang-Han Hsu[11], Guoqing Chang[1], Hsin Lin[11], Peisen Li[12], Kesong Yang[4,5], Zhenyu Wang[2,13,14,*], Tian Jiang[2,15], Xiao Renshaw Wang[1,16,*]

[1]*Division of Physics and Applied Physics, School of Physical and Mathematical Sciences, Nanyang Technological University, 21 Nanyang Link, 637371, Singapore*
[2]*State Key Laboratory of High Performance Computing, College of Computer Science and Technology, National University of Defense Technology, Changsha 410073, P. R. China*
[3]*Physics Laboratory, Industrial Training Center, Shenzhen Polytechnic, Shenzhen, Guangdong 518055, P. R. China*
[4]*Department of NanoEngineering and Program of Chemical Engineering, University of California San Diego, La Jolla, California 92093-0448, USA*
[5]*Program of Materials Science and Engineering, University of California San Diego, La Jolla, California 92093-0418, USA*
[6]*Institutes of Physical Science and Information Technology, Anhui University, Hefei 230601, P. R. China*
[7]*Fert Beijing Institute, MIIT Key Laboratory of Spintronics, School of Integrated Circuit Science and Engineering, Beihang University, Beijing 100191, P. R. China.*
[8]*Beihang-Goertek Joint Microelectronics Institute, Qingdao Research Institute, Beihang University, Qingdao 266000, P. R. China.*
[9]*Shanghai Key Laboratory of Special Artificial Microstructure Materials and Technology & School of Physics Science and Engineering, Tongji University, Shanghai 200092, P. R. China*
[10]*College of Advanced Interdisciplinary Studies, National University of Defense Technology, Changsha 410073, P. R. China*
[11]*Insitute of Physics, Academia Sinica, Taipei 115229, Taiwan*
[12]*College of Intelligence Science and Technology, National University of Defense Technology, Changsha 410073, P. R. China*
[13]*National Innovation Institute of Defense Technology, Academy of Military Sciences PLA China, Beijing 100010, P. R. China*
[14]*Beijing Academy of Quantum Information Sciences, Beijing 100084, P. R. China*
[15]*Beijing Institute for Advanced Study, National University of Defense Technology, Changsha 410073, P. R. China*
[16]*School of Electrical and Electronic Engineering, Nanyang Technological University, 50 Nanyang Ave, 639798, Singapore*

† These authors contributed equally
* Email: wangzy@baqis.ac.cn; renshaw@ntu.edu.sg

## Abstract

MnBi$_2$Te$_4$ (MBT) is the first intrinsic magnetic topological insulator with the interaction of spin-momentum locked surface electrons and intrinsic magnetism, and it exhibits novel magnetic and topological phenomena. Recent studies suggested that the interaction of electrons and magnetism can be affected by the Mn-doped Bi$_2$Te$_3$ phase at the surface due to inevitable structural defects. Here we report an observation of nonreciprocal transport, *i.e.* current-direction-dependent resistance, in a bilayer composed of antiferromagnetic MBT and nonmagnetic Pt. The emergence of the nonreciprocal response below the Néel temperature confirms a correlation between nonreciprocity and intrinsic magnetism in the surface state of MBT. The angular dependence of the nonreciprocal transport indicates that nonreciprocal response originates from the asymmetry scattering of electrons at the surface of MBT mediated by magnon. Our work provides an insight into nonreciprocity arising from the correlation between magnetism and Dirac surface electrons in intrinsic magnetic topological insulators.

## Introduction

MnBi$_2$Te$_4$ (MBT) is the first discovered intrinsic magnetic topological insulator, and it immediately sparked a surge of research interest in novel topological properties[1–3]. MBT possesses a layered stoichiometric antiferromagnetic (AFM) material with each septuple-layer (SL) formed in the sequence of Te-Bi-Te-Mn-Te-Bi-Te. The magnetic structure of MBT is A-type with intralayer ferromagnetic (FM) and interlayer AFM couplings[4,5]. Therefore, the ground magnetic state of MBT can be categorized into two types, namely FM with odd-numbered SLs and AFM with even-numbered SLs. Especially, the intrinsic magnetism interacts with the spin-momentum locked conduction electrons, resulting in a mass gap in the Dirac surface state[6]. By tuning the Fermi level into the opening gap, the quantum anomalous Hall effect and layer Hall effect have been realized in MBT[7–9]. When the Fermi level is located away from the gap, MBT may bring about rich spintronic functions, such as unidirectional magnetoresistance and the dissipative process by magnon[10,11]. Moreover, the recent observations of structural properties, *i.e.* surface collapse and reconstruction[12], temperature-independent bandgap[1,13] and even the gapless surface Dirac cone[14], revealed that magnetic and topological properties of MBT are sensitive to the surface structure. Inevitable Mn-doped $Bi_2Te_3$ phase exists in the MBT, which originates from the Te vacancy[12] or the FM island on some areas[15], can induce the magnetic moment at the surface[13] and weaken the interaction between magnetism and topological surface state.

Characterized by the current-direction-dependent electrical resistance, nonreciprocal transport[16,17] in quantum materials lacking inversion symmetry has triggered an outburst of investigations on various symmetry-related phenomena[18–21], *e.g.* ferroelectricity, noncentrosymmetric superconductivity and rectification effect. When further breaking the time-reversal symmetry, the nonreciprocal transport can be greatly enhanced[22,23] and provides insights into exotic properties[24–26], such as magneto-chiral anisotropy and bilinear magnetoresistance. Recently, nonreciprocal transport measurements were conducted to probe the asymmetry of materials, including the asymmetry of scattering processes by magnetic quasiparticles[21,27,28] and the asymmetry of electronic band structures with spin-splitting[11,22,25]. Investigating AFM MBT by the nonreciprocal transport measurement expects to reveal the intricate surface property. First, the spatial inversion symmetry is only broken at the surface of the MBT but preserved in the bulk[29,30], so that the nonreciprocal transport may only appear at the surface/interface of MBT. Second, the even-numbered MBT expects to be AFM with fully compensated magnetization, and the effect of uncompensated magnetic moment at the surface can be magnified. Therefore, the nonreciprocal transport may provide insight into the interaction of surface Dirac electrons and defect-induced magnetic moment at the surface[25].

In this work, we report the observation of nonreciprocal transport in a bilayer composed of AFM MBT layer and nonmagnetic (NM) Pt layer. The Pt capping layer is used to (i) protect the MBT layer from degradation, (ii) enhance the nonreciprocal transport by magnifying the asymmetry between top and bottom surfaces, and (iii) realize the current-induced magnetization switching in the intrinsic magnetic topological insulator for providing an alternative route of controlling the resistance. Nonreciprocal transport occurs when the temperature is below Néel temperature, $T_N$, revealing the association of nonreciprocity with inherent magnetism at the MBT surface. We show that both longitudinal and transverse nonreciprocal responses vary as a function of the angle between a current and in-plane magnetic field with a fixed π/2 offset. We elaborate the offset by the magnon-mediated scattering of electrons in the spin-momentum-locked surface state with uncompensated magnetization at the surface. Last, we show that uncompensated magnetization can be electrically switched with a current density of 9.26 x $10^{11}$ A $m^{-2}$.

## Results and discussion

We fabricated a Hall bar device with a bilayer structure composed of even-numbered SLs MBT and Pt capping layers on sapphire substrates. The single-crystalline MBT layer was grown using a molecular beam epitaxy (MBE) technique, which was considered as one of the best techniques of sample growth for investigating the surface properties[31–33]. The crystalline and electronic structures were characterized by reflection high energy electron diffractometer (RHEED), X-ray diffraction (XRD), low energy electron diffractometer (LEED), angle-resolved photoemission spectroscopy (ARPES) and scanning transmission electron microscope (STEM) (Fig. S1-S4). The Pt capping layers were deposited with a nominal thickness of 6 nm by magnetron sputtering. Figure 1a shows the schematic of the MBT/Pt bilayer and an optical image (Fig. 1b) of a representative Hall bar device. In our study, the polar angles in *yz*-, *xy*- and *xz*- planes are defined as *ξ*, *φ* and *θ*, respectively.

Figure 1c shows the longitudinal resistance, $R_{xx}$, as a function of an out-of-plane magnetic field in 12 SLs MBT/Pt bilayer at temperatures ranging from 2 to 25 K. The black dash lines in Fig. 1b shows that magnetoresistance, which is defined as MR = [$R_{xx}(H)$-$R_{xx}(0)$]/$R_{xx}(0)$, experiences two transitions at two critical fields. The two critical fields are denoted as $\mu_0 H_{c1}$ and $\mu_0 H_{c2}$ for the critical field and saturation field of spin-flop transition[34], respectively. As the magnetic field increases, the spin configuration evolves from A-type AFM to canted-AFM (c-AFM) at $\mu_0 H_{c1}$, and from c-AFM to FM at $\mu_0 H_{c2}$[1,35]. Figure 1d shows the diagram of the critical fields for different spin configurations of the MBT as a function of temperature. The two critical magnetic fields decrease as the temperature increases and become zero at a temperature higher than the $T_N$ of ~ 25 K, suggesting the phase transition from AFM to paramagnetic order[36].

Figure 1e shows transverse resistance, $R_{xy}$, in an out-of-plane magnetic field up to 8 T at 2 K. A weak anomalous Hall effect confirms the uncompensated magnetization in the MBT, indicating the uncompensated magnetic moment at the surface of MBT induced by inevitable Mn-doped $Bi_2Te_3$ phase. The nonsquare hysteresis loop observed in pure MBT thin film (Fig. S14) further confirmed the existence of the Mn-doped $Bi_2Te_3$ phase at the MBT surface. The black arrows show $\mu_0H_{c1}$ and $\mu_0H_{c2}$, which are consistent with the MR results (Fig. 1b). The disordered surfaces with experimentally observable anomalous Hall effect occur in a series of bilayers with different thicknesses of the MBT layer (Fig. S5). Figure 1f shows the temperature-dependent $R_{xx}$ of MBT layer, Pt layer and MBT/Pt bilayer. The MBT/Pt bilayer exhibits a comparable $R_{xx}$ to the Pt layer, whereas the $R_{xx}$ of MBT layer is three orders of magnitude larger than that of the Pt layer. Therefore, we confirmed the applied current mainly passing through the Pt layer. We also performed the magnetotransport measurement in the MBT thin film without capping Pt layer (Fig. S13). By comparing the magnetotransport data in MBT/Pt bilayer with the same thickness of MBT layer (Fig. S15), we confirmed that the response from the interface of MBT/Pt bilayer can be well detected.

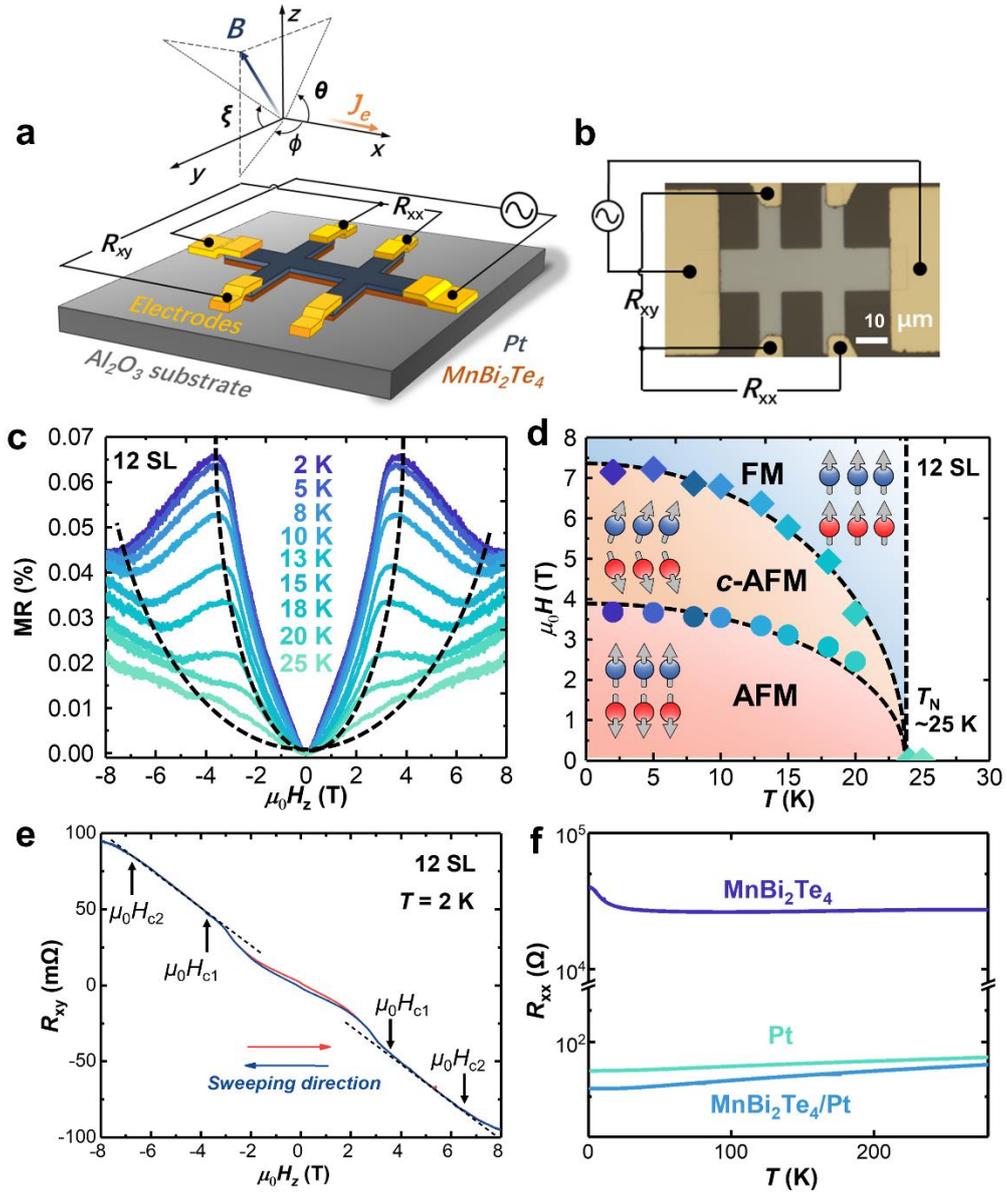

Figure 1. Magnetotransport characterization of 12 SLs MBT/Pt bilayer. (a) Measurement configuration of the MBT/Pt bilayer device. The polar angles in the *yz*-, *xy*-, *xz*-planes are defined as $\xi$, $\phi$ and $\theta$, respectively. (b) Optical image of a Hall bar device of a 12 SLs MBT/Pt bilayer. The scale bar is 10 μm. (c) Temperature-dependent MR of 12 SLs MBT/Pt bilayer at 2 to 25 K in an out-of-plane magnetic field up to 8 T. The dashed line shows the trend of the two critical magnetic fields, $\mu_0 H_{c1}$ and $\mu_0 H_{c2}$, at different temperatures. (d) Spin configuration of 12 SLs MBT/Pt bilayer as functions of temperature and external magnetic field. (e) The $R_{xy}$ of 12 SLs MBT/Pt bilayer measured at 2 K in an out-of-plane magnetic field up to 8 T. The black arrows also denote the $\mu_0 H_{c1}$ and $\mu_0 H_{c2}$. (f) Temperature-dependent $R_{xx}$ of MBT/Pt bilayer, MBT layer and Pt layer.

To suppress the bulk contribution in the conductance, reducing the bulk/surface ratio is necessary[37,38]. Hence, the measurements were performed in the 4 SLs MBT/Pt bilayer throughout the experiment unless otherwise stated. The $T_N$ ~ 21 K of 4 SLs MBT was determined through the critical power-law (Fig. S6). By applying an ac current $I(\omega) = I\sin(\omega t)^{39,40}$, the electrical voltage can be expressed as[17,41] $V = R_0 I(1 + \gamma BI)$, where $\gamma$ denotes the coefficient of nonlinear nonreciprocal transport. The term, $\gamma BI^2$, indicates that the nonreciprocal voltage depends on both current and magnetic field direction. Current-dependent nonreciprocal response is observed to be proportional to the square of applied current (Fig. S7). Therefore, nonreciprocal transport can be characterized by measuring the first ($V^{1\omega}$) and second harmonic voltage ($V^{2\omega}$). According to the direction of voltage, *i.e.* longitudinal and transverse voltage, we name the first (second) harmonic voltage as $V_{xx}^{1\omega}$ ($V_{xx}^{2\omega}$) and $V_{xy}^{1\omega}$ ($V_{xy}^{2\omega}$), respectively. Figure 2a exhibits the $V_{xy}^{2\omega}$ as a function of the *x*-direction in-plane magnetic field at 2 K. The $V_{xy}^{2\omega}$ as a function of the in-plane magnetic field can be separated into two regions, namely (i) low-field region and (ii) high-field region. In the low-field region, the nonreciprocal response scales linearly with the applied magnetic field. In the high-field region, the nonreciprocal response decreases with the magnetic field increases. This peculiar trend is a signature of asymmetric scattering process-induced nonreciprocity, which was also observed in other systems[11,42]. Figure 2b shows the magnitude of $V_{xy}^{2\omega}$ as a function of temperature under a fixed 2 T *x*-direction magnetic field. Notably, the $V_{xy}^{2\omega}$ gradually decreases as the temperature increases, and disappears when the temperature is higher than $T_N$ ~ 21 K, confirming the correlation of nonreciprocity and intrinsic magnetism in the MBT.

Figure 2c,d shows the magnetic field-dependent $V_{xy}^{1\omega}$ and $V_{xy}^{2\omega}$ within an in-plane magnetic field up to 0.5 T at 2 K. The $\mu_0 H_x$ (Fig. 2c) and $\mu_0 H_y$ (Fig. 2d) are in-plane magnetic fields with directions parallel and orthogonal to the applied current, respectively. The $V_{xy}^{1\omega}$ and $V_{xy}^{2\omega}$ shows parabolic and linear dependencies to the magnetic fields, respectively, which is similar to the behavior of $V^{2\omega}$ in conventional NM/FM heterostructures[43,44]. However, the magnitude of the $V_{xy}^{2\omega}$ under a *y*-direction magnetic field is negligible, which is in stark contrast to the conventional NM/FM heterostructure with a comparable thickness of Pt layer[44,45]. Hence, there shall be a different mechanism dominating the observed nonreciprocal response in the MBT/Pt bilayer.

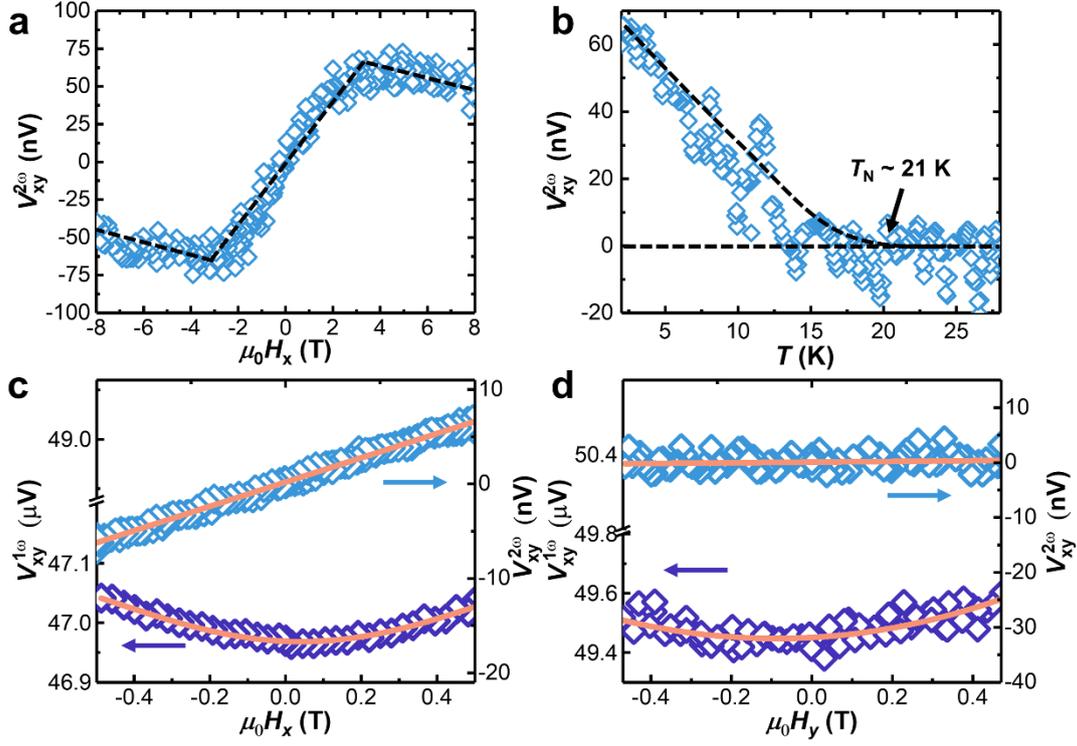

Figure 2. Nonreciprocal transport measurement of the 4 SLs MBT/Pt bilayer measured at 2 K. (a) $V_{xy}^{2\omega}$, as the function of *x*-direction magnetic field up to 8 T. (b) Temperature-dependent $V_{xy}^{2\omega}$ under the *x*-direction magnetic field whose magnitude was fixed at 2 T. The $V_{xy}^{2\omega}$ becomes negligible when the temperature is above $T_N \sim 21$ K. $V_{xy}^{1\omega}$ (purple diamond) and $V_{xy}^{2\omega}$ (blue diamond) are detected under the magnetic fields applied along *x*- (c) and *y*-directions (d), respectively. The data collected after +*M* initial magnetization are fitted using parabolic and linear functions (orange lines).

To investigate the origin of the nonreciprocal transport, we measured the angular dependent $V_{xx}^{2\omega}(\varphi, \theta, \xi)$ and $V_{xy}^{2\omega}(\varphi, \theta, \xi,)$ in *xy*-, *xz*- and *yz*-planes at 2 K (Fig. 3). The magnetic fields were identically set to 2 T to exclude the effect of multidomain formation[46]. Figure 3a,d shows the $V_{xx}^{2\omega}$ and $V_{xy}^{2\omega}$ as a function of $\varphi$ under an in-plane magnetic field. Notably, a π/2 angle offset between the anisotropic $V_{xx}^{2\omega}$ and $V_{xy}^{2\omega}$ was observed, which is identical to the nonreciprocal transport in the three-dimensional topological insulator and the magnet/topological insulator heterostructure[26,37,42]. The similar magnitudes of $V_{xx}^{2\omega}$ and $V_{xy}^{2\omega}$ indicate that the two nonreciprocal transport share the same microscopic origin[42]. Moreover, we believe that the π/2 angle offset in our magnetic MBT is due to an asymmetric scattering process of Dirac surface electrons mediated by magnon[11]. Figure 3g shows the schematic of gapped Dirac-like dispersion of the surface state in MBT[10], and the Fermi level $E_F$ is located in the conduction band which is electron-doped[1,13]. Figures 3h and 3i show the schematics of the asymmetric scattering processes at the perspective of the surface Fermi contour. The current direction is along the *x*-direction, and the magnetic fields are along the *x*- or *y*-direction. We chose four representative positions (*A*, *B*, *C* and *D*) based on the applied current (fixed along *x*-direction) and magnetic field (either *x*- or *y*-direction) directions. Because of the spin-momentum locking in MTI, the spin angular momentum, *S*, of electrons at the positions are locked, and their directions are shown in the schematics. In consideration of the quantization direction of MBT's magnetization, the angular momentum of the magnon is +1. We first discuss an example where the in-plane magnetic field and current are both along the *x*-direction (Fig. 3h). When an electron with *S*=-1/2 (at position *D*) is backscattered to the *S*=+1/2 (at position *B*), a magnon is emitted, based on the conservation of angular momentum. Similarly, the electron scattering from position *B* to *D* leads to magnon absorption. When switching the external magnetic field from *x*- to *y*-direction, similar emission and absorption of magnon during the electron's scattering between position *A* and *C* are expected. Because the scattering processes are nonequivalent, the relaxation times of electrons is asymmetric, leading to the different distributions of electrons at position *A* and *C*. Therefore, the asymmetry of electron distribution generates the nonreciprocal transport with the π/2 angle offset between anisotropic $V_{xy}^{2\omega}$ and $V_{xx}^{2\omega}$ under an in-plane magnetic field. The inhomogeneities at the MBT surface lead to the uncompensation of magnetic moment. And the local magnetic moment interacts with the surface Dirac electrons to produce the observed nonreciprocal transport.

We further discuss the possible contribution of the NM layer to the nonreciprocal response. The nonmagnetic layer as a source of damping-like and field-like torques has been widely studied in conventional NM/FM heterostructure[47–49]. However, if the $V^{2\omega}$ mainly originates from the magnetization oscillation due to the damping-like torque, the $V_{xx}^{2\omega}$ in *xz*-plane and $V_{xy}^{2\omega}$ in *yz*-plane is expected to be significant[50–52], in contrast to our experimental observation in Fig. 3b,f. In the scenario dominated by a field-like torque, the effective field exerting along y-direction gives rise to $V_{xy}^{2\omega}$ in yz-plane, which is also inconsistent with our observations (Fig. 3b). This in turn suggest the contribution of magnetization oscillation from the NM layer is negligible. In addition, NM systems with strong Rashba spin-orbit interaction can also produce nonreciprocal response[27]. However, the nonreciprocal resistance expects to linearly scale with the magnetic field, which is against our magnetic field dependence (Fig. 2a). We, therefore, confirm that the main contribution of observed nonreciprocal response originates from the asymmetric scattering of electrons by magnons.

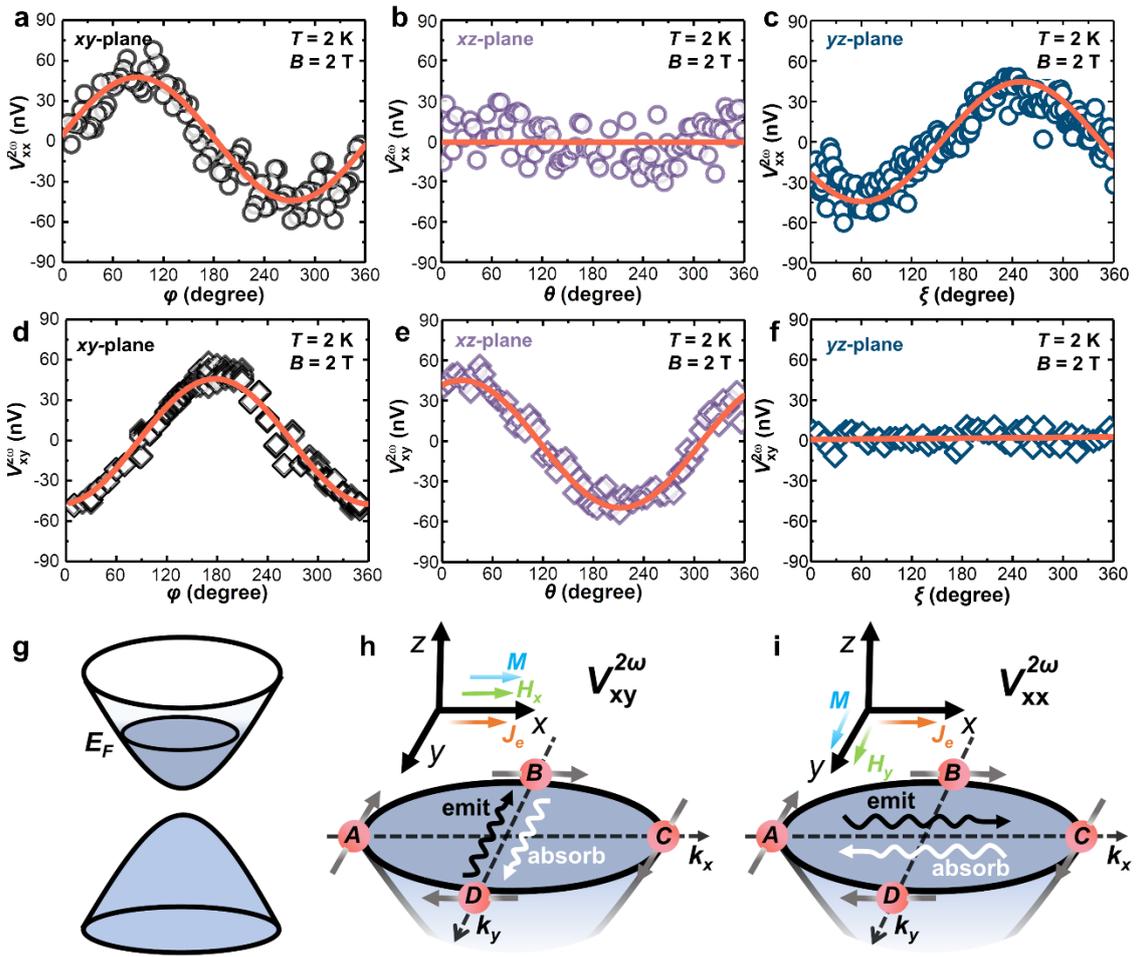

Figure 3. Angular dependent $V_{xx}^{2\omega}$ and $V_{xy}^{2\omega}$ in the 4 SLs MBT/Pt bilayer at 2 K. The $V_{xx}^{2\omega}$ and $V_{xy}^{2\omega}$ were collected as functions of rotation angle in the xy-plane (a,d), xz-plane (b, e), and yz-plane (c,f). The magnetic field was set to 2 T. (g) Gapped Dirac-like dispersion of the surface state in the MBT. Fermi level $E_F$ is located away from the gap due to the electron-doped nature of MBT (h, i) Schematic perspective view of the Fermi contour of surface state in the MBT under x- (h) and y-direction (i) magnetic fields. The asymmetric scattering of spin-polarized surface Dirac electrons is mediated by magnon emission (black arrow) and absorption (white arrow). The magnon emission and absorption occur between electrons at positions B and D under an

*x*-direction magnetic field (h), and between electrons at positions *A* and *C* under a *y*-direction magnetic field (i).

Last, electrically switching of uncompensated magnetization was achieved in the MBT/Pt bilayer. Figure 4a shows the schematic of current-induced magnetization switching in the MBT/Pt bilayer. By injecting charge current into the heavy metal under a small in-plane magnetic field $H_x$, the generated spin-orbit effective fields force the uncompensated magnetization of MBT switching the direction. A 10 ms write-current pulse up to 15 mA was applied along *x*-direction, followed by a read current 500 uA to measure the $R_{xy}$. Remarkably, the current-induced magnetization switching was realized with a small bias field applied along *x*-direction after positive polarization along +*z*-direction. Figure 4b shows the symmetric bipolar switching[48] at 2 K under a ±0.1 T bias magnetic field applied along the *x*-direction. When the magnetic field was applied along +*x*-direction, we observed an anticlockwise hysteresis as a function of pulsed current. In this hysteresis, high and low $R_{xy}$ states can be obtained after a positively- and negatively- pulsed current was applied. These two states (high and low $R_{xy}$) represent +*z*- and -*z*-direction of net magnetization, confirming the feasibility of electrically controllable magnetization at the surface of MBT. As we reversed the magnetic field direction, the hysteresis became clockwise, and the switching process was reversed accordingly. The relatively small ratio of reversed magnetization is proposed to the strong pinning effect from the bulk AFM. Figure 4c summarizes the critical current of reversing magnetization, $J_s$, as a function of the *x*-direction bias magnetic field. With the assist of a larger external magnetic field, the required spin-orbit effective field to switch the magnetization direction becomes smaller, so that the $J_s$ decreases. Figure 4d shows the temperature-dependent $J_s$ with a fixed bias field $H_x$ = 100 mT from 2 to 15 K. The $J_s$ decreases with temperature increases, due to the smaller net magnetization of the MBT layer at high temperature. Additionally, we confirmed that the influence of Joule heating and thermoelectric effect can be excluded (Fig. S11 and S12), and the electrically switched magnetization is repeatable in different devices (Fig. S10).

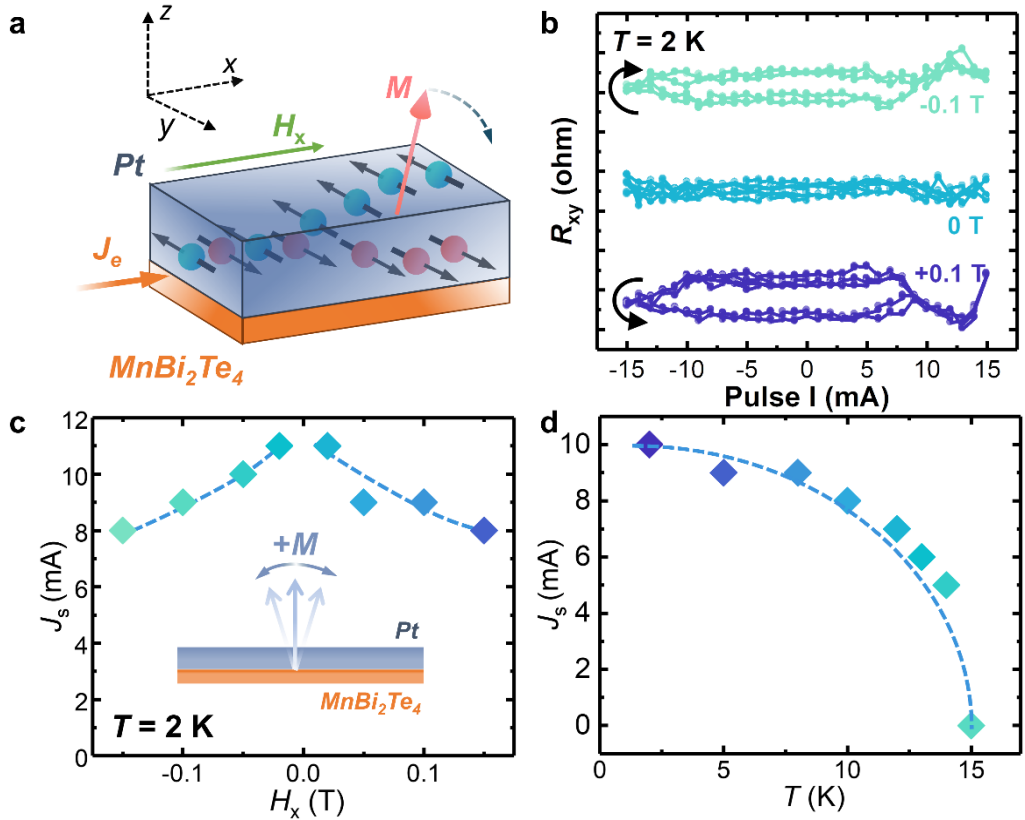

Figure 4. The electrical manipulation of magnetic order in the 4 SLs MBT/Pt bilayer. (a) Schematic diagram of magnetization reversing in MBT/Pt bilayer. Bypassing the current $J_e$ along $x$-direction with the assist of bias magnetic field ($H_x$). The switching of uncompensated magnetization in the MBT layer is determined by measuring the change of $R_{xy}$. (b) Bias-field-dependent write-read loops at 2 K. Scale bar is 0.1 mΩ. (c) Bias field-dependent critical current of magnetization switching at 2 K. (d) Temperature-dependent critical current of magnetization switching under a small bias field $H_x$ = 100 mT at temperatures from 2 to 15 K.

## Conclusion

In summary, we have demonstrated the nonreciprocal transport in the MBT/Pt bilayer. The emergence of nonreciprocity when the temperature is below $T_N$ reveals the correlation between nonreciprocal response and magnetism of MBT. The angular-dependent nonreciprocal transport under an in-plane magnetic field confirms the asymmetry of the magnon-mediated scattering process of electrons in the spin-momentum-locked surface state of MBT. The asymmetric relaxation times of electrons lead to the inequivalent electron distribution and thus the occurrence of nonreciprocal response. The unavoidable Mn-doped $Bi_2Te_3$ phase owing to the imperfect surface in the MBT induces the uncompensated magnetic moment. We further showed that the uncompensated magnetization can be electrically manipulated in the 4 SLs MBT/Pt heterostructure. Our results provide an insight into the surface transport properties and an alternative route of controlling resistance in the intrinsic magnetic topological insulator.

## Methods

**Material preparation**

High-quality $MnBi_2Te_4$ (MBT) films were grown on $Al_2O_3$(0001) substrates in a molecular beam epitaxy system with a base pressure of < $1\times10^{-10}$ mBar. Prior to the growth, the sapphire substrates were annealed at 750 °C for 2 h until the chamber vacuum reached the base pressure. High-purity Bi (99.9999%), Te (99.9999%) and Mn (99.99%) was evaporated from standard Knudsen cells. Other growth details and the characterizations of crystalline and electronic structures of the MBT films can be found in the Supplementary Information.

**Device fabrication**

The MBE-growth MBT films were transferred into a magnetron sputtering chamber for Pt deposition via an inert-gas-protected transfer chamber. The Pt films with a uniform thickness of ~6 nm were deposited on the MBT film by the magnetic sputtering. The MBT/Pt bilayers were then patterned into a Hall-bar structure with channel dimensions of 20×80 µm$^2$ by UV lithography. The etching was performed by Ar gas to remove the excess. Last, the Ti/Au (10/90 nm) electrodes were patterned in a second UV lithography process and deposited by e-beam evaporation.

**Electrical characterizations**

The electrical contacts were made by ultrasonically bonding Al wires onto the six electrodes on the Hall-bar devices. Transport measurements were conducted in an Oxford TeslatronPT cryostat with the temperature ranging from 300 to 2 K and the magnetic field up to 8 T. Keithley source meters (Keithley 6221 and 2182A) and lock-in amplifiers (Stanford SR830) were employed for the magneto-transport and nonreciprocal transport measurements. The 21 Hz a.c. read current with constant amplitudes of 500 uA was applied on the MBT/Pt Hall-bar devices for the magneto-transport measurements and nonreciprocal transport measurements. In the electrical manipulation characterizations, we first applied a pulse current along the *x*-direction with a 10 ms duration. Afterward, a constant 500 uA read current with a frequency of 21 Hz was used to measure the $R_{xy}$. The device's temperature under different amplitudes of pulse current was constantly monitored and evaluated by the corresponding $R_{xx}$.

## Supplementary information

More material preparation, sample structural characterization of $MnBi_2Te_4$ thin films; magnetotransport characterization of $MnBi_2Te_4$/Pt bilayer and $MnBi_2Te_4$ thin films.

## Acknowledgement

X.R.W. acknowledges supports from the Nanyang Assistant Professorship grant from Nanyang Technological University, Academic Research Fund Tier 2 (Grant No. MOE-T2EP50120-0006 MOE-T2EP50220-0016) and Tier 3 (Grant No. MOE2018-T3-1-002) from Singapore Ministry of Education, the Singapore National Research Foundation (NRF) under the competitive Research Programs (CRP Grant No. NRF-CRP21-2018-0003), and Agency for Science, Technology and Research (A*STAR) under its AME IRG grant (Project No. A20E5c0094). G.C. acknowledges supports from the Nanyang Assistant Professorship grant from Nanyang Technological University. Z.W. acknowledges supports from the General Program of Beijing Academy of Quantum Information Sciences (Project Y18G17), and the National Natural Science Foundation of China (Grant No. 12004434).

## Author contribution

C.Y. performed the electrical measurements with the help of K.H. and A.J.Y.. Z.W. and X.X. grew and characterized the MBT thin films with the help of M.T., T.Z., P.L. and T.J.. W.L.

fabricated the devices. S.J. and K.Y. provided the theoretical calculation. C.Y. and X.R.W. draft the paper and revised the paper with the help of D.Z., X.Q., C-H.H., G.C. and H.L.. All authors contributed to the scientific discussion. X.R.W. designed and directed the study.